\documentclass[utf8]{FrontiersinHarvard} 

\usepackage{url,hyperref,lineno,microtype,subcaption}
\usepackage[onehalfspacing]{setspace}

\def\keyFont{\fontsize{8}{11}\helveticabold }
\def\firstAuthorLast{Sima Roy {et~al.}} 
\def\Authors{Sima Roy\,$^{1}$ and Amar P. Misra\,$^{1,*}$}
\def\Address{$^{1}$Department of Mathematics, Siksha Bhavana, Visva-Bharati University, Santiniketan-731 235,  India  }
\def\corrAuthor{Corresponding Author}
\def\corrEmail{apmisra@visva-bharati.ac.in}

\begin{document}
\onecolumn
\firstpage{1}

\title[Stability of electromagnetic solitons]{Electromagnetic solitons and their stability in relativistic degenerate dense plasmas with two electron species} 

\author[\firstAuthorLast ]{\Authors} 
\address{} 
\correspondance{} 

\extraAuth{}

\maketitle

\begin{abstract}

The evolution of electromagnetic (EM) solitons  due to nonlinear coupling of circularly polarized intense laser pulses with low-frequency electron-acoustic perturbations is studied in relativistic degenerate dense astrophysical plasmas with two groups of electrons: a sparse population of classical relativistic electrons   and a dense population of relativistic degenerate electron gas. Different forms of localized stationary solutions are obtained and their properties are analyzed. Using the Vakhitov-Kolokolov stability criterion, the conditions for the existence and stability of a moving EM soliton are also studied.   It is noted that the stable and unstable regions shift around the plane of soliton eigenfrequency and the soliton velocity due to the effects of relativistic degeneracy, the fraction of classical to degenerate electrons and the   EM   wave frequency. Furthermore, while the standing solitons exhibit stable profiles for a longer time, the moving solitons, however, can be stable or unstable depending on the degree of electron degeneracy, the soliton eigenfrequency and the soliton velocity. The latter with an enhanced value can eventually lead to a soliton collapse.     The results should be useful  for understanding the formation of solitons in the coupling of   highly intense laser pulses with slow response of  degenerate dense plasmas in the next generation laser-plasma interaction experiments  as well as the novel features of $x$-ray and $\gamma$-ray pulses that originate from compact astrophysical objects.
\section{}
\tiny
 \keyFont{ \section{Keywords:} Electromagnetic soliton, Electron-acoustic wave, Degenerate plasma, Two-temperature electrons, Nonlinear Schr{\"o}dinger equation, Stability of solitons, Ponderomotive force} 
\end{abstract}

\section{Introduction}\label{sec-intro}
The nonlinear  propagation of intense electromagnetic (EM) waves in plasmas is typically associated with a wide variety of interesting   nonlinear phenomena, such as the generation of wakefields  \citep{roy2019}, parametric instabilities \citep{quesnel1997,barr1999,gleixner2020}, harmonic generation \citep{mori1993,shen1995}, self-focusing of wave envelopes \citep{esarey1997}, generation of intense electric and magnetic fields \citep{borghesi2002,wagner2004}, and  localization of EM waves as solitons \citep{farina2001,sundar2011,roy2019}. One  particular class  of waves that caught attention in the context of laser-plasma interactions in relativistic regimes is  the EM solitons. Many years back,  in $1956$, Akhiezer and Polovin  proposed  a set of relativistic electron fluid equations coupled to the Maxwell equations to model  the  interactions of intense EM waves with plasmas and found exact nonlinear wave solutions (solitons)   in relativistic plasmas. Such EM  solitons are  high-frequency laser pulses that are  typically modulated by   low-frequency plasma density perturbations and that propagate without any diffraction or spreading out of waves.  They appear in the context of various plasma environments and have potential applications, e.g.,  in   laser fusion and plasma-based particle accelerators \citep{sen1994,esirkepov2002,farina2005}.  Relativistic EM solitons have been observed in two-dimensional (2D) and three-dimensional (3D) particle-in-cell (PIC) simulations \citep{esirkepov2002,sentoku1999,bulanov1999} and also detected in experiments using   proton imaging techniques \citep{borghesi2002,borghesibulanov2002}. It has also been shown   \citep{esirkepov2002,bulanov1999} that  nearly twenty five to forty  percent of the laser pulse energy goes into the generation of EM solitons or soliton-like structures which may persist as appreciable signatures in the forms of modulated pulses in the radiation spectra (ranging from radio to $\gamma$-rays) that emanate from compact astrophysical objects. 
\par
The interaction of highly intense laser pulse or EM waves with ponderomotive force driven slow response of plasma density perturbations has been known to be one of the most powerful mechanisms for the generation of wakefields or the formation of EM solitons and hence for particle acceleration at relativistic speed in high intensity electromagnetic fields.  Although in high density regimes  the Fermi pressure seems to be the dominant effect,  the main dominating force in such nonlinear interactions is typically the EM wave ponderomotive force. It has been seen that the degeneracy Fermi pressure mainly contributes to the wave dispersion and hence plays key roles for  the transition from wakefield generation to the soliton formation [See, e.g., \citep{roy2019}   and references therein]. There are some other effects such as spin-orbit interaction \citep{morandi2014,asenjo2012,asenjo2011}, the Darwin term \citep{asenjo2012,asenjo2011}, and pair-production \citep{hebenstreit2011}  which may, however, become important in some other contexts than the present theory.  It is to be noted that  the mechanism of soliton formation considered here may  not be directly related to the quantum electrodynamics  in which the  Schwinger limit is applicable. The electrons achieve relativistic speed by the ponderomotive force of the highly intense EM fields, and high density degenerate plasmas can thus be generated simultaneously, e.g.,  with the production of $X$-rays and $\gamma$-rays  in the environments of compact astrophysical objects like white dwarfs. 
\par
The formation of envelope solitons in the interactions of  circularly polarized EM waves  with cold plasmas have been studied by Kozlov \textit{et al}. \citep{kozlov1979}. They employed the quasineutral approximations to establish the existence of small-amplitude localized solutions in the form of drifting solitons. A subsequent numerical investigation has been  focused on EM solitons of relativistic amplitude  \citep{kozlov1979}. Mima et al. \citep{mima1986} studied  the propagation characteristics of solitary waves  and predicted as possible   charged particle  acceleration mechanism. Furthermore,  Kaw et al. \citep{kaw1992} solved a  set of coupled nonlinear equations for modulated light pulses and electron plasma waves, and discussed the nonlinear relationship among the group velocity, amplitude, and  the frequency of   envelope solitons.    Farina and Bulanov \citep{farina2001} investigated the influence of ion motion on relativistic solitonic structures and they showed that the amplitude of moving solitons can be limited by this influence. In an another investigation and in a weakly relativistic regime, the nonlinear propagation of one-dimensional weakly nonlinear solitary waves in cold plasmas was studied using the reductive perturbation technique by Kuehl and Zhang \citep{kuehl1993}.   The nonlinear theory of weakly relativistic circularly polarized EM pulses in   warm plasmas  in the forms of bright and dark solitons was considered by Poornakala \textit{et al}  \citep{poornakala2002}  in different parameter regimes. The one-dimensional (1D) dynamics of   EM solitons in relativistic   electron–ion plasmas   has  also been studied  \citep{lontano2003,lontano2002}. Furthermore,   the existence of single-  and multiple-peak solitary structures and their stability, mutual interaction and propagation in   inhomogeneous cold plasmas were examined by Saxena \textit{et al}. \citep{saxena2006}. Recently, using the Vakhitov-Kolokolov criterion, the stability and dynamical evolution of linearly polarized  EM solitons were studied in the framework of  a generalized nonlinear Schr{\"o}dinger (GNLS) equation  in relativistic   degenerate dense plasmas  \citep{roy2020}. However, the present work considers the interactions of circularly polarized EM waves with two groups of electrons (one classical relativistic hot   electron species with a small concentration and the bulk relativistic degenerate electron gas).  While the two-electron species in plasmas  support low-frequency electron-acoustic waves \cite{misra2021}, the  single species electrons with stationary background of ions as in \citep{roy2020} are   rather relevant  for   high-frequency Langmuir waves. So, the evolution equations and the results, to be obtained in the present study,  will be significantly different from those of the work of \citep{roy2020}. 
 In an another work, the nonlinear coupling of EM waves and low-frequency electron-acoustic density fluctuations was considered by Shatashvili \textit{et al.} \citep{shatashvili2020} in relativistic astrophysical plasmas with two-temperature electrons. They reported that the modulated EM waves can propagate in the subsonic or supersonic regimes. However, the existence criterion, the dynamical evolution and stability of  EM solitons were not studied in this investigation. 
\par 
The aim of this work is to   advance the previous theory  \citep{shatashvili2020} of EM waves in relativistic degenerate dense plasmas and to report the detailed analysis for the existence of different kinds of localized soliton solutions, their stability analysis using the Vakhitov-Kolokolov criterion,  as well as their dynamical evolution by a simulation approach.  It is found that the parameter domains for the stable and unstable regions predicted in the linear analysis well agree with the simulation results.   
The motivation for studying a two-temperature electron plasma is due to fact that although there is no direct observational evidence for the existence of  two groups of electrons in relativistic degenerate regimes,  based on the information available from the theories and some relevant observations,  it is expected that such highly relativistic  degenerate astrophysical plasmas coexisting with classical relativistic hot electron flow  can exist, e.g.,  during the formation of relativistic jets  due to accretion-induced collapsing of white dwarfs into black holes \citep{begelman1984,kryvdyk1999,kryvdyk2007}.   The relativistic  dense plasmas where the  background distribution of electrons deviates from the thermodynamic equilibrium can also appear  in the context of laser produced plasmas   or ion beam driven plasmas \citep{hau2011,gibbon2005}. In such cases, the system energy mostly flows into the electrons, thereby generating fully degenerate electrons with long tails or   partially degenerate electrons with high temperature tails, and allowing the division of  background electrons  with two different temperatures.  The excitation of low-frequency electron-acoustic waves in these systems  are believed to   play key  roles in the nonlinear  wave coupling and the formation of coherent structures like solitons.
\section{Basic equations} \label{sec-BEs}
The nonlinear coupling of high-frequency circularly polarized intense EM waves and slow plasma response of low-frequency electron-acoustic perturbations in   unmagnetized plasmas composed of a dense relativistic degenerate electron gas (with number density $n_d\equiv n_{d0}+\delta n_d$), a sparse population of nondegenerate classical electrons (with number density $n_\text{cl}\equiv n_{cl0}+\delta n_\text{cl}$), and immobile ions is governed by the following Zakharov-like equations \citep{shatashvili2020}.
\begin{equation}
2i\left(\frac{\partial}{\partial t} + v_g \frac{\partial}{\partial z}\right)A + b_{0}\frac{\partial^{2} A}{\partial z^{2}}+b_{1}NA  + b_{2}|A|^2 A=0, \label{eq1}
\end{equation}
\begin{equation}
\left(\frac{\partial^{2}}{\partial t^{2}}-\frac{\partial^{2}}{\partial z^{2}}\right)N = -3b_{3}\frac{\partial^{2}|A|^2}{\partial z^{2}}, \label{eq2}
\end{equation}
where $N \equiv \delta n_d/n_{d0}$ is the dimensionless degenerate electron density perturbation and $A \equiv eA/mc^{2}$ is the dimensionless EM wave vector potential with  $e$ denoting the elementary charge, $m$   the electron mass, and $c$   the speed of light in vacuum. In Eqs. \eqref{eq1} and \eqref{eq2}, the time and space coordinates are normalized according to $t\rightarrow t\omega_{0}$ and $z\rightarrow z\omega_{0}/c_{s}$, where $\omega_0~(k_0)$ is the EM wave frequency (wave number) given by the high-frequency dispersion relation: $\omega_0^2=c^2k_0^2+\Omega_d^2+\alpha\omega_{ed}^2$ and $c_s$ is the electron-acoustic speed, defined by, $c_s^2=(1/3)\alpha c^2 R_0^2/\sqrt{1+R_0^2}$. Here, $\omega_{ed}\equiv \sqrt{4\pi n_{d0}e^2/m}$ is the plasma oscillation frequency of degenerate electrons, $\Omega^2_d=\omega^2_{ed}/\sqrt{1+R_0^2}$, $\alpha=n_{cl0}/n_{d0}~(\ll1)$ is the ratio between the  equilibrium number densities of classical and degenerate electrons, and $R_0=\hbar(3\pi^2n_{d0})^{1/3}/mc$ measures the degree of electron degeneracy, i.e., $R_0\ll1~(\gg1)$ corresponds to weakly relativistic (ultra-relativistic) degenerate plasmas.   Also, $v_g$ is the  EM wave group velocity normalized by   $c_{s}$, $b_{0}=\left(dv_{g}/dk_0\right)\omega_{0}/c_{s}^{2}$ represents the group velocity dispersion,    $b_{1}=\left(\omega_{ed}/\omega_{0}\right)^2\left(1-\kappa^{2}/\sqrt{1+R_{0}^{2}}\right)$ is associated with the nonlocal nonlinearity, $b_{2}=\left(\omega_{ed}/\omega_{0}\right)^2\left(\alpha+\kappa^{2}/(1+R_{0}^{2})^{3/2}\right)$ is the coefficient of the cubic nonlinear term, and $b_{3}= \left(\sqrt{1+R_{0}^{2}}-\kappa^2\right)/R_{0}^{2}$ with $2/3<\kappa^{2}\equiv1-R_{0}^{2}/3(1+R_{0}^{2})<1$ corresponds to the ponderomotive nonlinearity.  
\par 
The model equations \eqref{eq1} and \eqref{eq2} have been derived using a set of relativistic fluid equations coupled to the Maxwell equations on several  assumptions, namely,  the thermodynamic temperature of classical electrons is much lower than the Fermi temperature,   the characteristic wave frequency is much higher than the de Broglie frequency, and the EM pump wave is weakly relativistic. Furthermore, it has been assumed that the two electron species have different effective masses $(m_d>m_\text{cl})$ in which   the effective mass of classical electrons $(m_\text{cl})$ is determined by their thermodynamic temperature while that of degenerate electrons $(m_\text{d})$ is determined by their number density    \cite{shatashvili2020}.   The model was essentially derived to explore novel nonlinear coupling of high-frequency EM waves with low-frequency electron-acoustic waves and the modulational instability induced by the new physics originating due to the flow of classical relativistic electrons in relativistic degenerate plasmas \cite{shatashvili2020}.  Such plasmas, coexisting with a classical hot accumulating astrophysical flow, are interesting and important plasma ingredients in the environments of white dwarf stars, e.g., during the relativistic jet formation from collapsing white dwarfs to black holes \citep{begelman1984,kryvdyk1999,kryvdyk2007}. The model can similarly be used to explore the new particle-acceleration mechanism via the formation of wakefield generation and EM solitons in which the degeneracy pressure can play a decisive role for the transition from wakefield generation to soliton formation \citep{roy2019}.       
\par 
It is to be noted that although there are similar works in the literature along the lines of the present study, no previous results in the literature can be recovered  in some particular limits, namely nonrelativistic plasma flow or weakly relativistic degeneracy  (classical) of the present model equations \eqref{eq1} and \eqref{eq2}.  However, there are some studies on electromagnetic envelope solitons in relativistic magnetized plasmas based on the  NLS formalism. For example, starting from a set of relativistic fluid equations coupled to the circularly polarized EM wave equation,  Borhanian \textit{et al.} \citep{borhanian2009}   studied the modulational instability and the evolution of circularly polarized EM wave envelopes in magnetized plasmas. Their formalism was based on a multiple scale perturbation approach to derive the NLS equation (without any nonlocal nonlinearity) which is distinctive from the present approach. Furthermore, the existence and the properties of standing high-frequency  EM solitons were studied  by Mikaberidze \textit{et al.} \citep{mikaberidze2015} in a fully degenerate overdense electron plasma  by considering a set of relativistic fluid and Maxwell equations.  Their evolution equations for traveling EM solitons are similar to Eqs. \eqref{eq16} and \eqref{eq17}  obtained in Sec.  \ref{sec-LSS}.  Since the formalism considered in these studies is different from the present  one, one cannot recover the previous results even in some particular cases as mentioned before. 
\section{Localized stationary soliton}\label{sec-LSS}
Before we investigate the conditions for the existence of EM solitons and their stability, we first look for some simple localized stationary soliton solutions that relativistic two-temperature plasmas can support and that are correlated with the plasma density depletion. So, looking for a localized stationary soliton solution of Eqs. \eqref{eq1} and \eqref{eq2}, we introduce
  the coordinate transformations $\xi=z-v_gt$, $\tau=t$, and   assume   the vector potential $A$ to be of the form $A=a(\xi)\exp(i\omega\tau)$ and $N\equiv N(\xi)$, where $\omega$ is the EM soliton eigenfrequency. Under these transformations Eqs.  \eqref{eq1} and \eqref{eq2} reduce to
\begin{equation} \label{eq16}
b_{0}\frac{d^{2} a}{d \xi^{2}}-2\omega a+b_{1}Na+b_{2}a^{3}=0,
\end{equation}
\begin{equation} \label{eq17}
(v_g^{2}-1)\frac{d^{2} N}{d\xi^{2}}= - 3b_{3}\frac{d^{2}a^{2}}{d\xi^{2}}.
\end{equation}
Solving Eq. \eqref{eq17} yields the following expression for the density perturbation.
\begin{equation}\label{eq18}
N=\frac{3b_{3}}{1-v_g^{2}}a^{2}.
\end{equation}
Here, we note that since $2/3<\kappa^2<1$, $b_3$ is always positive. Also,   from the high-frequency dispersion relation   and the expression for the electron-acoustic speed $c_s$  stated in Sec. \ref{sec-BEs},  it can be noted that  $v_g>1$ holds  (for which we have a density depression) in the moderate or weakly relativistic degenerate regime. However, $v_g<1$ holds (for which we have a density hump) for $R_0\gg1$, i.e., in the ultra-relativistic regime. Thus, it follows that the stationary localized EM solitons may exist  in a wide range of values of the degeneracy parameter $R_0$.   Since the density perturbation cannot vanish in the region of EM field localization, there is no possibility of the formation of cavitation at some point in the region. 
\par 
Next, eliminating $N$ from Eqs. \eqref{eq16} and \eqref{eq18}, we obtain the following nonlinear differential equation for the wave amplitude $a$.
\begin{equation} \label{eq19}
\frac{d^{2} a}{d \xi^{2}}- \mu a + f(a^2)a=0,
\end{equation} 
where $\mu=2\omega/b_0$ is the nonlinear frequency shift   measuring the inverse of the square of the characteristic width of the soliton and   $f$ is the nonlinear function of the wave amplitude, given by,
   \begin{equation}
   f(a^2)=\frac{1}{b_0}\left(b_{2}+\frac{3b_{1}b_{3}}{1-v_g^{2}}\right)a^2.
   \end{equation}
Clearly,  $|f|$ is an increasing function of $a$. However, considering $f$ as a function of $k_0$ (or $\omega_0$)  we note that while its absolute value increases in a small interval $0\lesssim k_0\lesssim0.1$, the same decreases in the other small sub-interval $0.1< k_0\lesssim0.3$  and tends to vanish for $k_0\gtrsim0.3$. In the latter, the localized soliton solution may not exist.       Furthermore, $|f|$ becomes higher  $(>1)$ with  higher  values of the degeneracy parameter $R_0\gtrsim30$ and with a small increment of  $\alpha$ and $\omega$.      Thus, there must be some parameter restrictions   for which the nonlinearity is not too high, i.e., $|f|\lesssim 1$ or a bit more   for the existence of EM solitons with amplitude $a\sim{ O}(1)$. 
\par 
In what follows, we integrate  Eq. \eqref{eq19} and use the boundary conditions, namely  $a \rightarrow 0$, $da/d\xi \rightarrow 0$, and $d^{2}a/d\xi^{2} \rightarrow 0$ as $\xi \rightarrow \pm \infty$ to obtain the following energy balance equation  for the motion of a pseudoparticle of unit mass in which   $a$ plays the role of a pseudo-coordinate and   $\xi$  that of the pseudo-time.
\begin{equation}  \label{eq 20}
\frac{1}{2}\left(\frac{da}{d\xi}\right)^{2} + V(a)=0,
\end{equation} 
where the pseudopotential $V(a)$ is given by
\begin{equation} \label{eq 21}
V(a)=\frac{1}{4b_{0}}\left(b_{2}+\frac{3b_{1}b_{3}}{1-v_g^{2}}\right)a^{4}-\frac{1}{2}\mu a^{2}.
\end{equation}
For the  existence of   finite-amplitude EM solitons,   the pseudopotential $V$ must satisfy the following conditions:
\begin{itemize}
\item[(i)]{$V(0)=V^{\prime}(0)=0$.}
\item[(ii)]{$V^{\prime\prime}(a)<0$ at $a=0$, so that the fixed point at the origin becomes unstable. }
\item[(iii)]{$V(a_{m}\neq0)=0$ and $V^{\prime}(a_{m})\gtrless0$ according to when the solitary waves are compressive (with positive potential, i.e., $a>0$) or rarefactive (with negative potential, i.e., $a<0$). Here, $a_{m}=\sqrt{4\omega/[b_{2}+3b_{1}b_{3}/(1-v_g^{2})]}$ represents the soliton amplitude.}
\end{itemize}
When the  above conditions are satisfied one can then integrate  Eq.  \eqref{eq 20} and use the boundary conditions stated before to find the following soliton solution.
\begin{equation} \label{eq 23}
a(\xi)=a_{m}~\text{sech} (\xi/\Delta),
\end{equation}
where    $\Delta=\sqrt{b_{0}/2\omega}$ is the width of the EM soliton. Since the amplitude $a_m$ is to be real, we must have either $v_g<1$ or $v_g>\sqrt{1+3b_{1}b_{3}/b_{2}}$. We note that  $v_g<1$ is satisfied either in the regime of $k_0\ll1$ for a moderate value of $R_0$ or in the regime $0\le k_0\lesssim 1$    with $R_0\gg1$.   Thus,   EM solitons in relativistic degenerate plasmas can travel as subsonic and supersonic waves \citep{shatashvili2020}. 
 \par
\begin{figure*}
\centering
   \includegraphics[width=6in, height=3in]{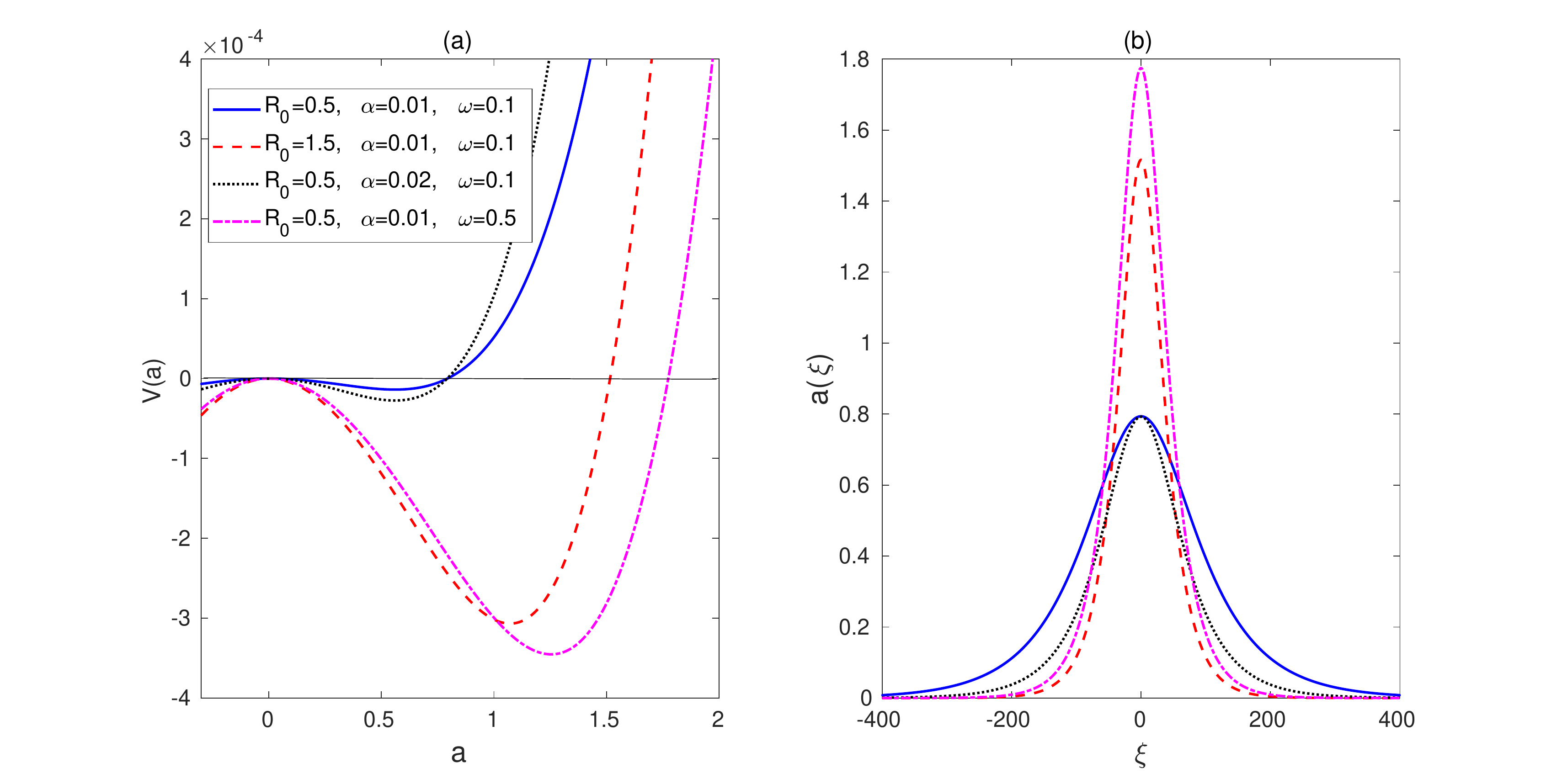}
   \caption{The Sagdeev potential $V(a)$ [subplot (a)]  and the corresponding soliton solutions $a(\xi)$ [subplot (b)] are shown for different values of  the parameters as in the legends  with a fixed value of $k_0=0.3$. }
\label{fig:sag-sol}
\end{figure*}

 
 We numerically verify the aforementioned conditions for the existence of EM solitons in different parameter regimes and plot the Sagdeev potential $V(a)$ against $a$ and the corresponding soliton profiles as shown in Fig. \ref{fig:sag-sol}. We note that the conditions for $V(a)$ are satisfied in a wide range of values of $R_0$ including $R_0\ll1$ and $R_0\gg1$ implying that EM solitons can exist  both  in the weakly relativistic and ultra-relativistic regimes. However, as noted before in highly degenerate regimes   $(R_0\gtrsim30)$, the nonlinear function $f$   tends to become much higher in magnitude. So, such highly degenerate regimes may not be admissible for the  existence of soliton  solution in the particular form. 
From the subplot (a) of Fig. \ref{fig:sag-sol} it is evident that for certain parameter values, $V(a)$ crosses the $a$-axis at $a=a_m$ and $V<0$ for $0<a<a_m$. Such $a_m$ is the amplitude of the soliton as is evident from the profiles   in subplot (b). The width of the soliton can also be obtained either using the relation $W=|a_m/V_\text{min}|$ from the profiles of $V(a)$ [subplot (a)]  or from the profiles  of $a(\xi)$ [subplot (b)]. Here,  $V_\text{min}$ represents the absolute minimum value of $V$. When    a value  of either the degeneracy parameter $R_0$ or the soliton frequency $\omega$ (or the wave number $k_0$) is increased, a significant increment (reduction) of the  soliton  amplitude (width) is noticed. However, the amplitude remains the same and the width gets reduced when a fraction of classical to degenerate electrons are slightly enhanced.  
\section{Stability of EM solitons} \label{sec-SEMs}
\begin{figure*}
\centering
\includegraphics[scale=0.36]{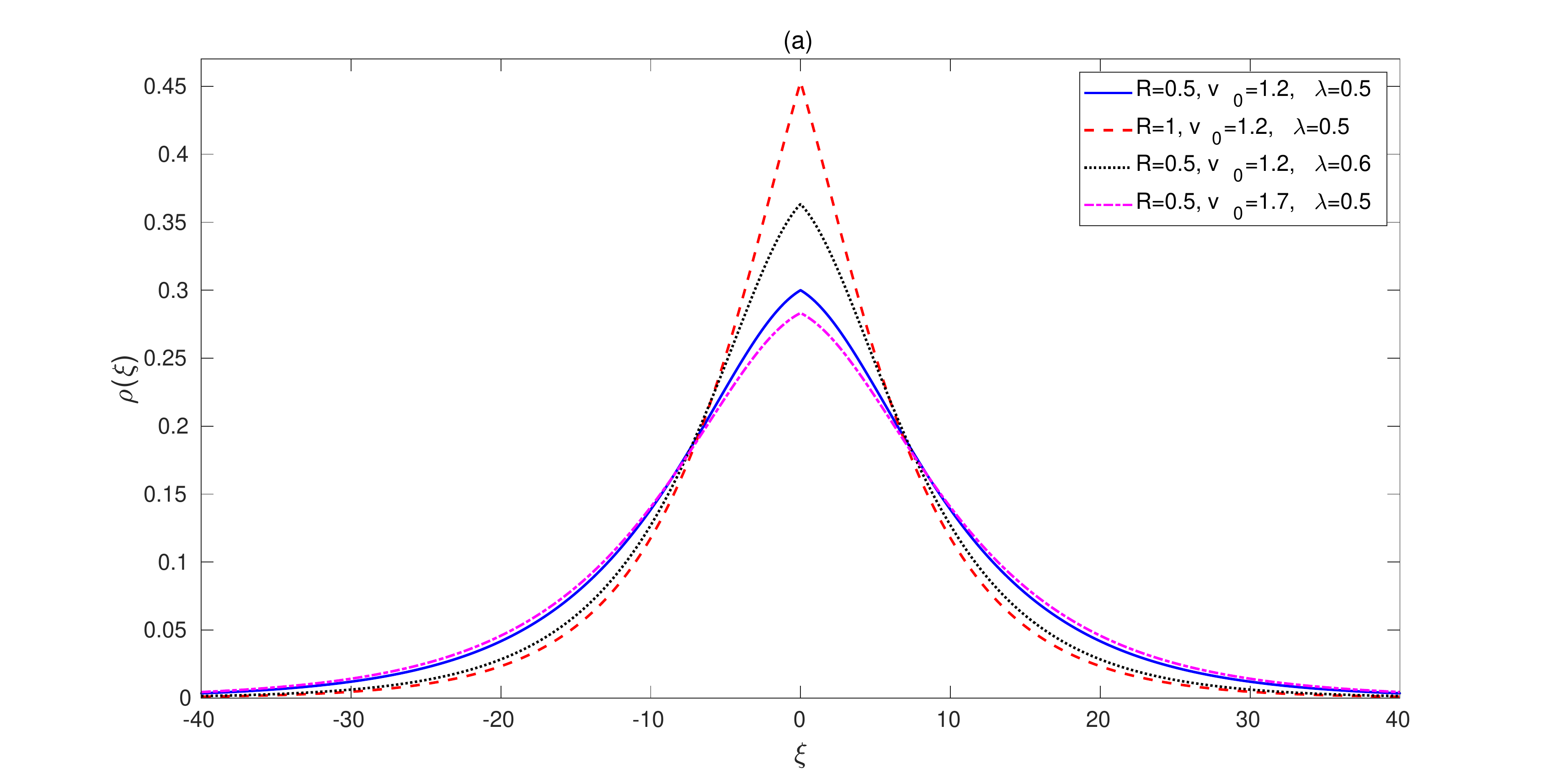}
\includegraphics[scale=0.36]{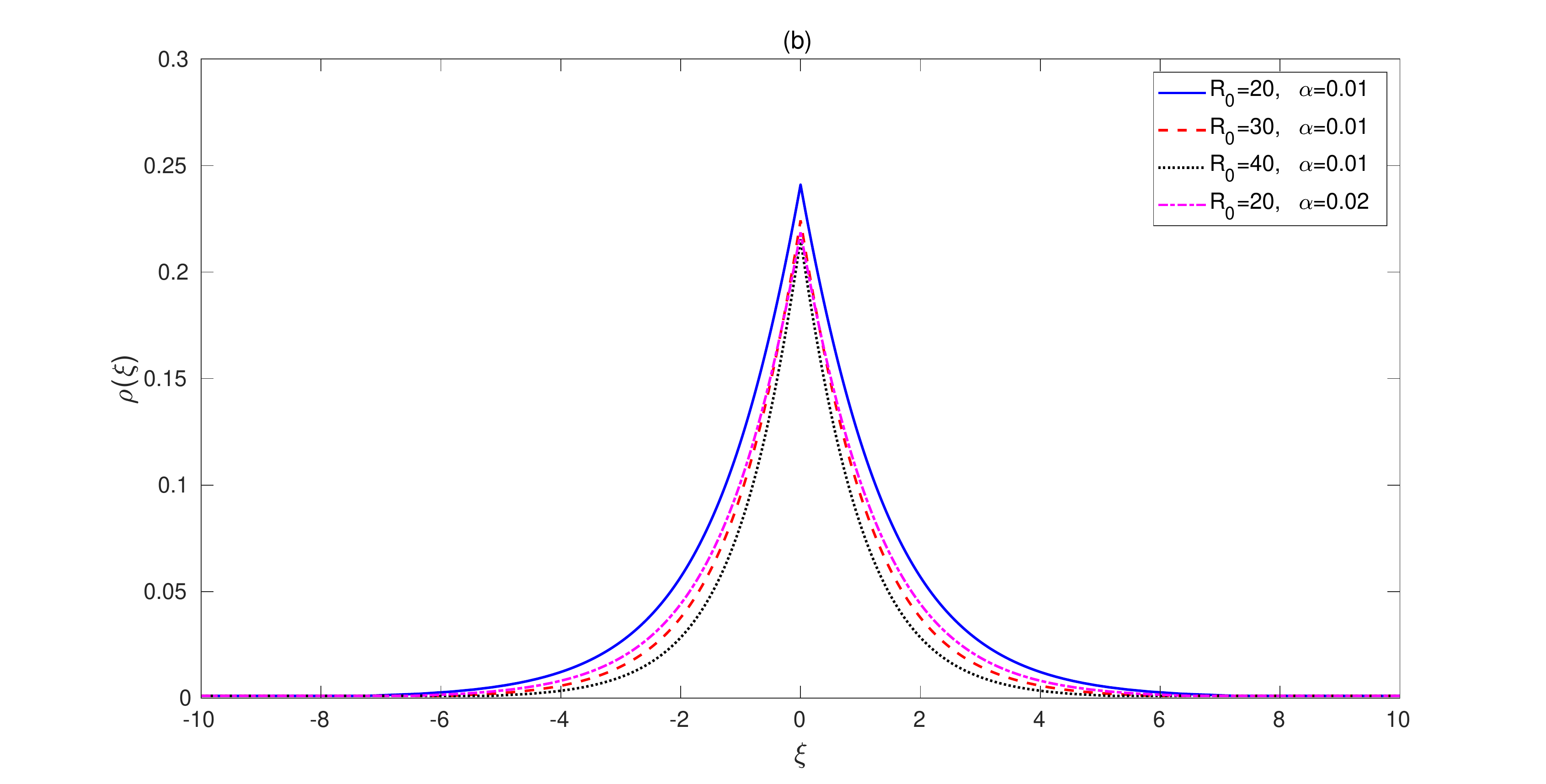}
\caption{Two branches of soliton solutions [Eq. \eqref{eq-soliton}], symmetric about the origin, are plotted against $\xi$  with a fixed $\alpha=0.01$ and different values of the  parameters as in the legends. While   subplot (a) is shown for a moderate value of $R_{0}$, subplot (b) is for relatively a higher value of $R_0$.   The soliton characteristics are significantly modified due to  different degrees of relativistic degeneracy.    }
\label{fig:soliton1}
\end{figure*}

In this section, we focus on the evolution of slowly varying weakly nonlinear small amplitude circularly polarized EM wave envelopes and their stability in relativistic degenerate dense plasmas. So, we introduce a slowly varying complex envelope  in the form:
\begin{equation} \label{eq-expn}
A=\frac{1}{2}(ae^{-it}+a^{*}e^{it}),~~
N=\frac{1}{2}(N_{2}e^{-i2t}+N_{2}^{*}e^{i2t}),
\end{equation}
where we have considered the odd harmonics (first order) for the vector potential $A$ and even harmonics (second order) for the density perturbation $N$. Since we are looking for an evolution equation for the wave potential $A$ of the NLS type, the contributions of  other harmonics may not be so important.   In Eq. \eqref{eq-expn}, the asterisk denotes the complex conjugate of the physical quantity.   Substituting the expansions of Eq. \eqref{eq-expn}   into Eq.    \eqref{eq2} and collecting the second harmonic terms $\sim \exp(-i2t)$,  we obtain  the following expression for $N_{2}$.
\begin{equation}
N_{2}=\frac{3}{8}b_{3}\frac{\partial^2 a^2}{\partial z^2}.\label{eq 4}
\end{equation}
Next,  substituting the expansions of Eq. \eqref{eq 4} into Eq.  \eqref{eq1} and collecting the first harmonic terms $\sim \exp(-it)$, we obtain the following   equation for the EM wave amplitude $a$.
\begin{equation}
i\left(\frac{\partial}{\partial t}+v_g\frac{\partial}{\partial z}\right)a+P\frac{\partial^2 a}{\partial z^{2}}+Q_1|a|^2a+Q_2 \frac{\partial^2 a^2}{\partial z^2}  a^{\ast} =0. \label{eq-nls0}
\end{equation}
Equation \eqref{eq-nls0} can further be reduced to the following form by applying the  transformations  $\xi=z-v_g t$ and $\tau=t$, i.e.,
\begin{equation}
i\frac{\partial a}{\partial \tau}+P\frac{\partial^2 a}{\partial \xi^{2}}+Q_1|a|^2a+Q_2 \frac{\partial^2 a^2}{\partial \xi^2}  a^{\ast} =0, \label{eq-nls1}
\end{equation}
where the coefficients of the dispersion $(P)$ and the local $(Q_1)$  and nonlocal $(Q_2)$ nonlinear coefficients are 
\begin{equation}
P=\frac{1}{2}b_0,~ Q_1=\frac{3}{8}b_2,~Q_2=\frac{3}{32}b_1b_3.
\end{equation}
Equation \eqref{eq-nls1} is in the form of a generalized nonlinear Schr{\"o}dinger (GNLS) equation in which the cubic (local) and derivative (nonlocal) nonlinearities are significantly modified by the effects of the relativistic degeneracy of dense electrons, percentage of nondegenerate electrons as well as the EM pump wave frequency.  Here, by the local nonlinearity  we mean the effect that occurs due to the interactions of different kinds of carrier harmonic modes (including self-interactions) at nonlinear regimes. On the other hand, the nonlocal nonlinearity appears due to the ponderomotive force of EM wave fields on the slow response of plasma density fluctuations.    In the limit of $R_0\ll1$, the nonlocal effect tends to become less significant and the local cubic nonlinearity prevails. However, as one gradually enters from the weakly to ultra-relativistic regimes with  $R_0\gg1$, the nonlocal term  dominates over the local nonlinearity in the domain of higher values of $k_0$. Thus, it becomes significant to study the evolution of localized wave envelopes and their stability especially in the relativistic degeneracy regime. So, we look for a localized stationary solution of Eq. \eqref{eq-nls1} in the form of a moving soliton (distinctive from that in Sec. \ref{sec-LSS}) as $a=\rho(\eta)\exp\left[i\theta(\eta)+i\lambda^{2}\tau\right]$ where $\eta=\xi-v_{0}\tau$ with $v_{0}$ denoting the soliton velocity (which, in general, is different from $v_g$) in the moving frame of reference.  Next, substituting this solution into Eq. \eqref{eq-nls1}, we obtain the following coupled equations for the soliton phase and the amplitude.
\begin{equation} \label{eq 7}
\beta\theta_{\eta\eta}\rho+\left(4\beta-2b_0\right)\rho_{\eta}\theta_{\eta}-2v_{0}\rho_{\eta}=0, 
\end{equation} 
\begin{equation} \label{eq 8}
\rho_{\eta\eta}+\frac{3}{8\beta}b_{1}b_{3} \rho \rho_{\eta}^{2}=\frac{3}{4\beta}\rho^{3}\left(b_{1}b_{3}\theta_{\eta}^{2}-b_{2}\right)+\frac{\rho}{\beta}(2\lambda^{2} -2\theta_{\eta}v_{0} +b_{0} \theta_{\eta}^{2}),
\end{equation}
where $\beta=b_{0}+(3/8)b_{1}b_{3}\rho^{2}$. Integrating Eq.  \eqref{eq 7} and using the boundary conditions, namely $\rho, \rho_{\eta}, \rho_{\eta\eta}\rightarrow 0$ as $\eta\rightarrow \pm\infty$, we obtain    $d\theta/d\eta\approx v_{0}/b_{0}$, whereas Eq. \eqref{eq 8} gives on integration
\begin{equation} \label{eq 9}
\rho_{\eta}^{2}=\frac{\rho^{2}}{b_{0}\beta}\left[2\lambda^{2}b_{0}-v_{0}^{2}-\frac{3}{8}b_{0}\left(b_{2}-b_{1}b_{3}\frac{v_{0}^{2}}{b_{o}^{2}}\right)\rho^{2} \right]. 
\end{equation}
Further integration of \eqref{eq 9} gives a soliton solution in the following implicit form.
\begin{equation} \label{eq-soliton}
\pm \eta=\sqrt{\frac{3b_{1}b_{3}\rho_{0}^{2}b_{0}}{8\left(2\lambda^{2}b_{0}-v_{0}^{2}\right)}}\tan^{-1}\left(\sqrt{\frac{8\beta/3b_{1}b_{3}}{\rho_{0}^{2}\left(1-\rho^{2}/\rho_{0}^{2}\right)}}\right)+\frac{(b_{0}/2)}{\sqrt{2\lambda^{2}b_{0}-v_{0}^{2}}}\ln\left|\frac{\sqrt{\beta}-\sqrt{b_{o}(1-\rho^{2}/\rho_{0}^{2})}}{\sqrt{\beta}+\sqrt{b_{o}\left(1-\rho^{2}/\rho_{0}^{2}\right)}}\right|,    
\end{equation}
where $\rho_0$ is the maximum amplitude of the   soliton, given by,
\begin{equation} \label{eq-rho0}
\rho_{0}^{2}=\frac{8}{3}\frac{\left(2\lambda^{2}b_{0}-v_{0}^{2}\right)b_0}{ b_{2}b_0^2-b_{1}b_{3}v_{0}^{2}}.
\end{equation}
From Eqs. \eqref{eq-soliton} and \eqref{eq-rho0} it follows that the  conditions for the existence of real   solutions require 
\begin{equation} \label{eq 14}
 2\lambda^{2}b_0-v_{0}^{2}>0,~~
 b_{2}b_0^2-b_{1}b_{3}v_{0}^{2}>0.
 \end{equation} 
 We will consider these conditions and discuss about the existence domains in more details shortly. In fact, Eq. \eqref{eq-soliton} describes two branches of soliton solutions which are symmetric about the origin and which, when combined together, form a complete soliton profile. The characteristics of these solitons are displayed in Fig. \ref{fig:soliton1} for different values of the parameters, namely the degeneracy parameter $R_0$, the soliton velocity $v_0$ and the eigenfrequency $\lambda$.  
 It is noted that in contrast to the effects of the soliton velocity,   i.e.,  by increasing values of which the soliton amplitude decreases but the width broadens, the effects of the electron degeneracy and the eigenfrequency are   to enhance the soliton amplitude but to reduce the width significantly as similar to those observed  in Fig. \ref{fig:sag-sol}.  The   percentage of classical electrons $\alpha$ has also the similar effects on the soliton profiles as in Fig. \ref{fig:sag-sol}. Physically, for the present model, the wave dispersion is provided by the degeneracy pressure of electrons  as well as by the   separation of charged particles (Poisson equation). However, as the values of the degeneracy parameter $R_0$ increase,   the  coefficients of the group velocity dispersion $P$ and the nonlocal nonlinearity (associated with the ponderomotive force) $Q_2$ tend  to decrease but those of the cubic nonlinearity $Q_1$ increase   and remain smaller than $Q_2$ in the regime $0<R_0\lesssim20$. As a result,   the increased wave energy from amplification is accommodated by an increase in soliton amplitude and a reduction in its width.  If the   density ratio $\alpha$ is slightly increased or if one considers the high- or ultra-relativistic degeneracy regimes, the ponderomotive nonlinearity tends to dominate  over the cubic nonlinearity which   results into a depletion of both the amplitude and width. However, some  other nonlinear phenomena including the soliton collapse can also take place which we will examine in the end of this section.
 \par 
 It is to be mentioned that the profiles in Fig. \ref{fig:soliton1}  are not exactly the  cusp solitons. The spiky shape appears due to numerical plots of the two functions with plus and minus signs in Eq.  \eqref{eq-soliton} separately, which seem to mismatch at the top of the curves.  The latter may be a sort of numerical error.  However, cusp solitons may appear in the context of NLS equations with saturating nonlinearities [See, e.g., \citep{wadati1980}].   
\par
\begin{figure*}
\centering
\includegraphics[scale=0.4]{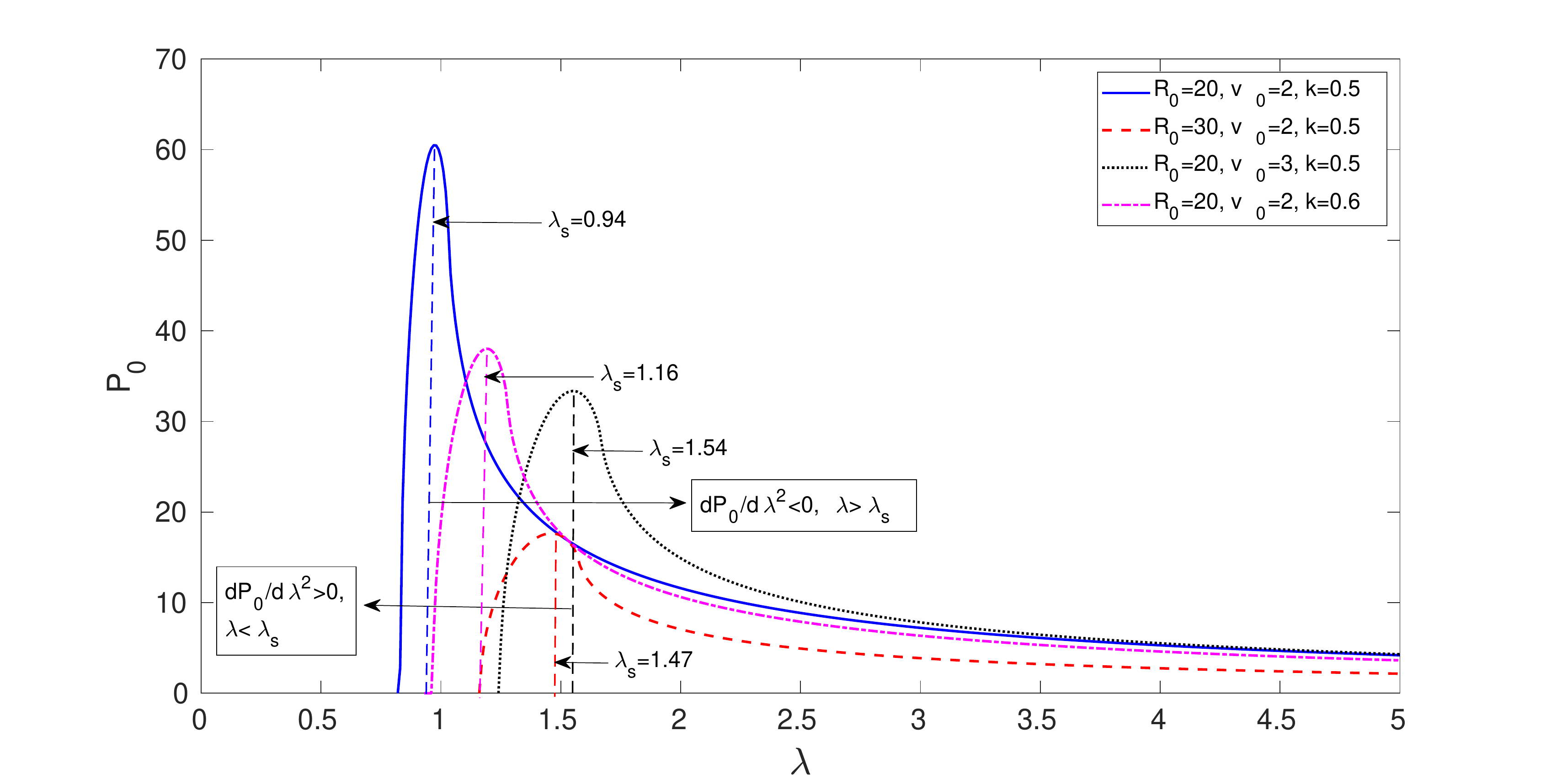}
\caption{The soliton photon number $P_0(\lambda)$ is plotted against the eigenfrequency $\lambda$ with a fixed $\alpha=0.01$ and for different values of the other parameters as in the legends. The solitons are stable (unstable)  in the regions $\lambda<\lambda_s,~dP_0/d\lambda^2>0$   ($\lambda>\lambda_s,~dP_0/d\lambda^2<0$).     }
\label{fig:photon}
\end{figure*}

 
It is now imperative to study  the stability of the moving   solitons   given by  Eq. \eqref{eq-soliton}. To this end, we use the Vakhitov-Kolokolov stability criterion \citep{vakhitov1973} according to which the solitons are said to be stable against a longitudial perturbation  if
\begin{equation}\label{eq 11} 
\frac{dP_{0}}{d\lambda^{2}}>0,
\end{equation}  
 where $P_{0}$ is the soliton photon number, defined by, 
\begin{equation}\label{eq 12}
P_0=\int|a|^{2} d\xi.
\end{equation}
The expression for $P_{0}(\lambda)$ can be obtained   by integrating Eq. \eqref{eq 9} with respect to $\eta$ with   the limits for $\rho$ as $0$ and $\rho_0$, and using Eq. \eqref{eq 12}      as
\begin{equation}\label{eq 13}
P_{0}(\lambda)=\frac{\sqrt{8}b_{0}}{\sqrt{3\left(b_{2}b_{0}^{2}-b_{1}b_{3}v_{0}^{2}\right)}}\times \left[\frac{\sqrt{2}\beta_{0}}{\sqrt{3b_{1}b_{3}}}\cot^{-1}\left(\sqrt{\frac{8b_{0}}{3b_{1}b_{3}\rho_{0}^{2}}}\right)-\rho_{0}\sqrt{\frac{b_{0}}{4}}\right],
\end{equation}
where $\beta_{0}=b_{0}+(3/8)b_{1}b_{3}\rho_{0}^{2}$ (the value of $\beta$ at $\rho=\rho_0$). Thus, according to the stability condition  [Eq. \eqref{eq 11}], the moving EM soliton \eqref{eq-soliton} is said to be stable in the region $\lambda<\lambda_{s}$ and unstable in the region $\lambda>\lambda_{s}$, where   $\lambda_{s}$ is some critical value of $\lambda$ at which   $P_{0}$ achieves a local maximum and beyond which $dP_{0}/d\lambda^{2}<0$. Since finding an analytic form of $\lambda_s$ is extremely difficult, we try to obtain its values by a numerical approach for different sets of values of the parameters. The profiles of $P_0(\lambda)$ and the corresponding values of $\lambda_s$ are exhibited in Fig.   \ref{fig:photon} with the variations of the degeneracy parameter $R_{0}$, the soliton velocity $v_{0}$ and the classical to degenerate  electron number density ratio $\alpha$. The stable and unstable regions can be predicted, respectively, with the conditions $\lambda<\lambda_s,~dP_0/d\lambda^2>0$ and $\lambda>\lambda_s,~dP_0/d\lambda^2<0$.   It is found that, an increase of each of $R_{0}$, $v_{0}$ and $k$ shifts the instability threshold $\lambda_{s}$ towards its higher values, implying that the stability domains $(\lambda<\lambda_s)$ are significantly enhanced. This means  that, in contrast to linearly polarized EM waves \citep{roy2020},   the circularly polarized EM solitons cab be stable in high density degenerate plasmas with two-temperature electrons.  
\begin{figure*}
\centering
\includegraphics[width=7in,height=3in]{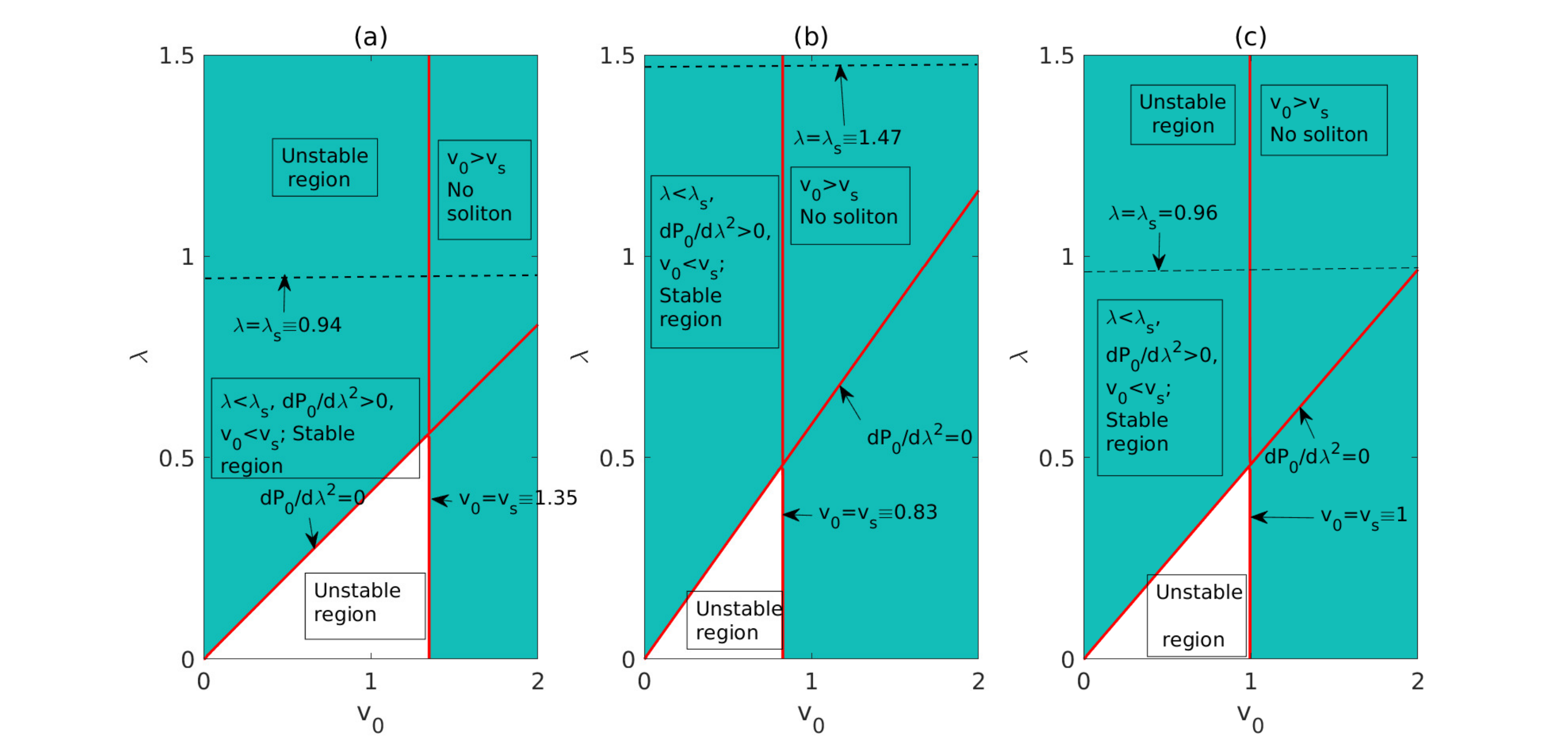}
\caption{The stable and unstable regions as well as the existence domains of  EM solitons are shown   in the $(v_{0},\lambda)$-plane for diferent values of the   parameters: (a) $R_0=20,~\alpha=0.01,~k_0=0.5$, (b)  $R_0=30,~\alpha=0.01,~k_0=0.5$, and    (c)  $R_0=20,~\alpha=0.01,~k_0=0.6$. }
\label{fig:region}
\end{figure*}
\par
Next, apart from the Vakhitov-Kolokolov stability criterion, we also examine if there be any   constraints on the parameters $\lambda$ and $v_{0}$ in order to have a real soliton solution \eqref{eq-soliton}. These constraints  together with the stability criterion stated before  can provide the existence as well as   the stable and unstable regions of EM solitons. In this context, we find the    limits for the parameters $\lambda$ and/or $v_{0}$ from Eq. \eqref{eq 14} as
 \begin{equation} \label{eq 15}
v_{0}<v_s\equiv  b_0 \min\left\{\sqrt{2}\lambda,\sqrt{b_{2}/b_{1}b_{3}}\right\}.
\end{equation} 
Thus, considering Eqs. \eqref{eq 14}, \eqref{eq 11},  and \eqref{eq 15}, we obtain the regions  for the existence of EM solitons and their   stability/instability   in the $(v_{0},\lambda)$-plane as shown in Fig. \ref{fig:region}.   Different critical values of $\lambda$ and $v_0$ are indicated by the text arrows and so are $v_0>v_s$ where no soliton solution exists. Thus, the existence regions together with the stable and unstable regions are those satisfy the conditions: $\lambda<\lambda_s,~v_0<v_s,~dP_0/d\lambda^2>0$ (Stable) and $\lambda<\lambda_s,~v_0<v_s,~dP_0/d\lambda^2<0$ (Unstable) as indicated in the subplots. We find that as the value of the degeneracy parameter $R_0$ increases, the stable/unstable regions shift towards unstable/stable ones with a significant reduction of $v_s$ but a significant increase   of $\lambda_s$. The wave number $k_0$ has also similar effects on the stable and unstable regions, however, a change in $\lambda_s$ is not so effective. Although in the region of $v_0>v_s$, no real analytic soliton solution [\textit{cf}. Eq. \eqref{eq-soliton}] exists, there may be some other numerical solution of Eq. \eqref{eq-nls1}. Such a discrepancy may occur due to an initial small phase difference between the approximate analytical and numerical solutions in the particular region. It is to be noted that the Vakhitov-Kolokolov criterion is suitable only for the linear stability of EM solitons that involve exponential growth or decay of wave modes. It cannot predict the nonlinear evolution of unstable envelope solitons or the stability of localized solutions that have arbitrary profiles. However, due to the presence of both the cubic and nonlocal nonlinearities, the GNLS equation can admit, apart from the envelope solitons, the soliton collapse or some other nonlinear features which   are out of scope of the present study.
 \par
\begin{figure*}
\centering
\includegraphics[width=7in,height=3in]{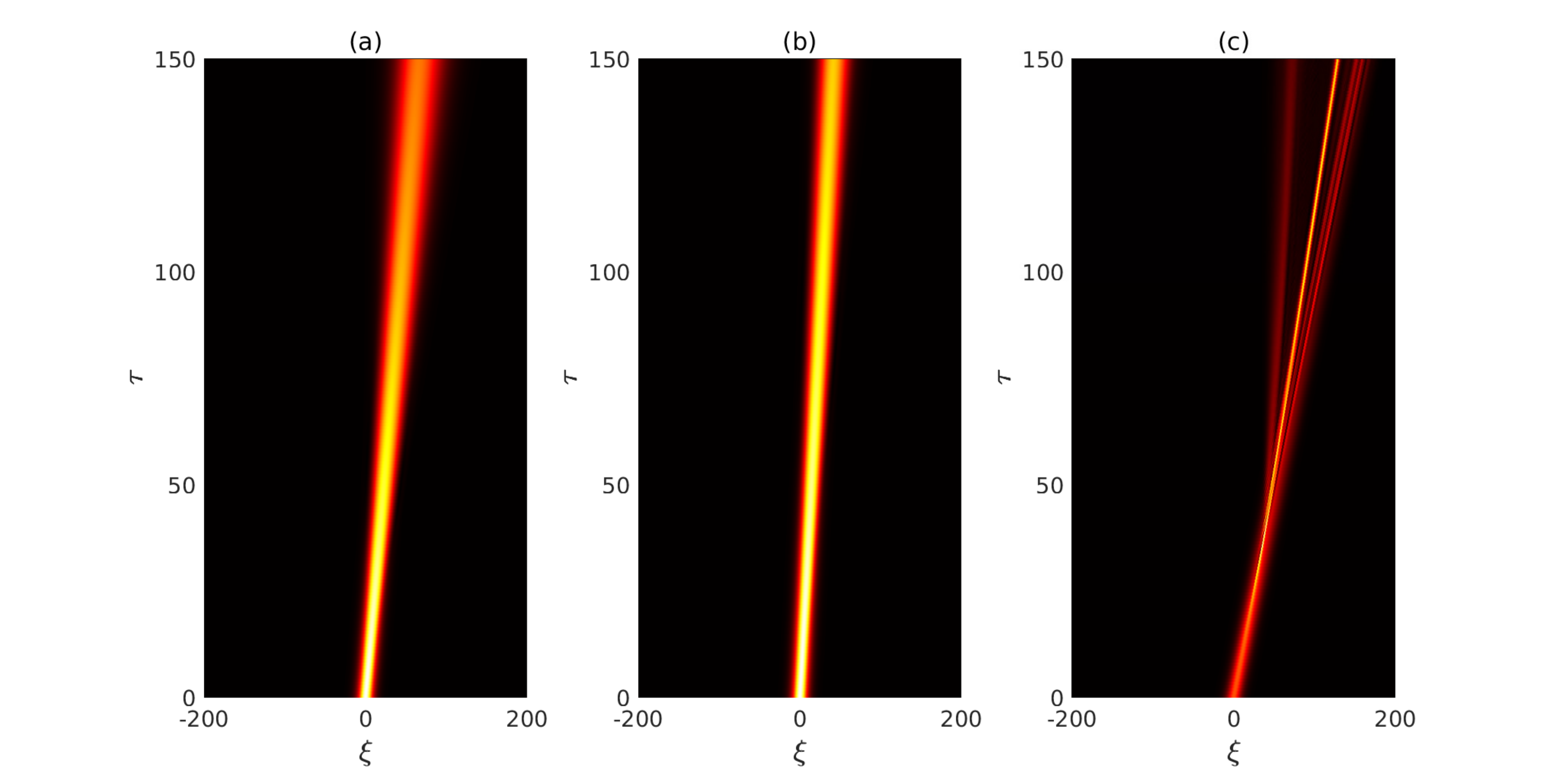}
\caption{Dynamical evolution of EM solitons [Numerical solution of Eq. \eqref{eq-nls1}] for different parameter values that correspond to stable [subplot (b)] and unstable [subplots (a) and (c)]  regions [\textit{cf}. Fig. \ref{fig:region}]. The parameter values for the subplots (a) to (c), respectively, are $R_0=20,~\lambda=1.5,~v_0=0.6,~\alpha=0.01,~k_0=0.5$; $R_0=30,~\lambda=0.2,~v_0=0.6,~\alpha=0.01,~k_0=0.5$; and $R_0=30,~\lambda=0.2,~v_0=2,~\alpha=0.01,~k_0=0.5$. }
\label{fig:soliton3}
\end{figure*}

 
Relying on the existence as well as the   stable and unstable regions of EM solitons obtained so far in the $(v_0,\lambda)$-plane, we now study the dynamical evolution of EM solitons by a numerical simulation approach. The aim is also to verify whether these regions indeed support the numerical soliton solutions.  So, we solve Eq. \eqref{eq-nls1} by using the Runge-Kutta scheme with a time step size $\Delta\tau=0.001$ and   the initial condition (at $\tau=0$): $a(\xi)\sim a_0\rm{sech}^2(\xi/8)\exp(-iv_0\xi)$. Figure \ref{fig:soliton3} shows the evolution of EM solitons after time $\tau=150$ with the spatial interval $-200\leq\xi\leq200$ and $1000$ grid points for different values of the parameters that fall in the existence and stable/ unstable regions. As an illustration,  we consider two unstable regions with (i) $R_0=20,~\lambda=1.5,~v_0=0.6,~\alpha=0.01,~k_0=0.5$ [subplot (a)] and (ii) $R_0=30,~\lambda=0.2,~v_0=5,~\alpha=0.01,~k_0=0.5$ [subplot (c)]  and a stable region with $R_0=30,~\lambda=0.2,~v_0=0.6,~\alpha=0.01,~k_0=0.5$ [subplot (b)].  It is seen that as time goes on, the initial profile tends to radiate and as the nonlinear and dispersion effects intervene the dynamics, the soliton  evolves into either a stable pulse or an unstable one. Furthermore, the standing soliton  with $v_0=0$, $R_0=20$, amplitude $a_0\sim0.7$,  $\lambda=0.2$, and photon number $P_0=1.64$ in the stable region oscillates around $\xi=0$ with a frequency close to the EM wave frequency and it remains stable for a longer time interval. However, as the velocity is increased, the moving soliton with $v_0=0.6$, $R_0=30$, amplitude $a_0\sim0.7$,  $\lambda=0.2$, and photon number $P_0=17.6$ in the stable region propagates towards the upstream region and it displays an increase of the wave  amplitude [subplot (b)]. Physically, an increase of the soliton velocity results into a reduction of the relative eigenfrequency $\Lambda\equiv\sqrt{\lambda^2-v_0^2/b_0}$ but an enhancement  of the photon number which, in turn, increases the soliton amplitude. However, as time progresses  the moving soliton mitigates to a stable structure.   
On the other hand, by increasing   the eigenfrequency $ \lambda=1.5$ but retaining the soliton speed at $v_0=0.6$ with a different value of $R_0=20$ [subplot (a)], we find that the photon number is reduced to $P_0=1.05$. As a result the soliton in the unstable region exhibits a decay of its amplitude  and as time elapses, it reaches towards a steady state stable soliton.  Next, further considering an increased value of the soliton velocity $(v_0=5)$, however,  retaining $R_0=30$ and  $\lambda=0.2$ that fall in the unstable region  we find that the soliton photon number increases and    the soliton  can no longer travel undistorted, i.e.,    its amplitude aperiodically grows and eventually leads to a collapse [subplot (c)].            
\section{Summary and conclusion}
We have studied the generation of EM solitons in the nonlinear interactions between a circularly polarized intense EM pulse and low-frequency electron-acoustic density fluctuations that are driven by the EM wave ponderomotive force in relativistic degenerate dense plasmas with two groups of electrons. The evolution of such solitons is described by a coupled set of nonlinear equations for the EM vector potential and the density fluctuations associated with the  slow plasma response \citep{shatashvili2020}. Stationary localized soliton solutions of the coupled equations are obtained and their characteristics are analyzed with the parameters that correspond to the relativistic degeneracy of electrons $(R_0)$, the fraction of classical to degenerate electrons $(\alpha)$ and the EM wave frequency $(\omega_0)$.
\par
Using the Vakhitov-Kolokolov stability criterion, we have also studied the existence conditions and performed a linear stability analysis of a  moving soliton whose evolution is governed by a GNLS equation with both the cubic (local) and derivative (nonlocal) nonlinearities.  Different stable and unstable regions are obtained which   shift around the $(v_0,\lambda)$-plane due to variations of the parameters $R_0$, $\alpha$, and $\omega_0$ that correspond to different physical regimes in astrophysical settings. Here, $v_0$ is the soliton velocity  and $\lambda$ is the eigenfrequency.   A direct numerical simulation of the generalized  GNLS  equation  reveals that the parameter domains for the stability and instability of EM solitons  well agree with those predicted using the Vakhitov-Kolokolov stability criterion. It is also found that an initially launched  moving soliton with an increased velocity in the instability domain   eventually collapses after a finite interval of time due to higher nonlocal nonlinear effects than the cubic nonlinearity.  
\par 
To conclude, it has been observed that relativistic high density degenerate plasmas deviating from thermodynamic equilibrium can appear   not only in the context of laser produced plasmas or beam driven plasmas but also in compact astrophysical objects like white dwarf stars, neutron stars \citep{hau2011,gibbon2005}. In these environments, since the system energy flows mostly into the electrons, they may appear either as  a group of partially degenerate electrons with high temperature tails or a group of relativistic  classical   and fully degenerate electrons. Such plasmas are known to support low-frequency electron-acoustic waves \citep{misra2021} which play key roles in the nonlinear wave dynamics.  Furthermore, these compact astrophysical objects emanate different EM radiation spectra ranging from radio to $\gamma$-rays. So, interactions of these intense pulses with high-density plasmas may give rise the formation of solitons and other coherent structures as localized bursts of $x$-rays and $\gamma$-rays. In this respect, the present theoretical results could be useful for understanding the characteristics of $x$-ray  and $\gamma$-ray pulses as well as for the next generation intense laser-solid density plasma interaction experiments.


\section*{Conflict of Interest Statement}

The authors declare that the research was conducted in the absence of any commercial or financial relationships that could be construed as a potential conflict of interest.

\section*{Author Contributions}

All authors  contributed to the manuscript equally.

\section*{Funding}
 A. P. Misra acknowledges support from Science and Engineering Research Board (SERB, Government of India) through a research project with sanction order no. CRG/2018/004475. 



\section*{Data Availability Statement}
 The original contributions presented in the study are included in the article. Further inquiries can be directed to the corresponding author.

\bibliographystyle{Frontiers-Harvard} 
\bibliography{References.bib}


\end{document}